\documentclass[aps,prl,twocolumn,showpacs,superscriptaddress,10pt,nofootinbib]
{revtex4-1}

\usepackage{graphicx}
\usepackage[ulem=normalem]{changes}

\def\ket#1{|#1\rangle}

\begin{document}

\title{Linking partial and quasi dynamical symmetries in rotational nuclei}
\author{C. Kremer}
\affiliation{Institut f\"ur Kernphysik, Technische Universit\"at Darmstadt, 
D-64289 Darmstadt, Germany}
\author{J. Beller}
\affiliation{Institut f\"ur Kernphysik, Technische Universit\"at Darmstadt, 
D-64289 Darmstadt, Germany}
\author{A. Leviatan}
\affiliation{Racah Institute of Physics, The Hebrew University, 
Jerusalem 91904, Israel}
\author{N. Pietralla}
\affiliation{Institut f\"ur Kernphysik, Technische Universit\"at Darmstadt, 
D-64289 Darmstadt, Germany}
\author{G. Rainovski}
\affiliation{Faculty of Physics, St. Kliment Ohridski University of Sofia, 
Sofia 1164, Bulgaria}
\author{R. Trippel}
\affiliation{Institut f\"ur Kernphysik, Technische Universit\"at Darmstadt, 
D-64289 Darmstadt, Germany}
\author{P. Van Isacker}
\affiliation{Grand Acc\'el\'erateur National d'Ions Lourds, CEA/DSM-CNRS/IN2P3,
B.P.~55027, F-14076 Caen Cedex 5, France}
\date{\today}
			

\begin{abstract}
\begin{description}
\item[Background] Quasi dynamical symmetries (QDS) and partial dynamical symmetries (PDS) play an important role in the understanding of complex systems. Up to now these symmetry concepts have been considered to be unrelated. \item[Purpose] Establish a link between PDS and QDS and find an emperical manifestation. \item[Methods] Quantum number fluctuations and the intrinsic state formalism are used within the framework of the interacting boson model of nuclei. \item[Results] A previously unrecognized region of the parameter space of the interacting boson model that has both O(6) PDS (purity) and SU(3) QDS (coherence) in the ground band is established. Many rare-earth nuclei approximately satisfying both symmetry requirements are identified. \item[Conclusions] PDS are more abundant than previously recognized and can lead to a QDS of an incompatible symmetry.
\end{description}
\end{abstract}	
	
	
\maketitle

Understanding the structure and dynamics of complex many-body systems 
can often be obtained from the observation and analysis of symmetries. 
Symmetry considerations are particularly significant for addressing 
a key question in such systems, namely, 
how do simple features emerge within a complicated environment.
A notable example is the collective behavior of nuclei
which stems from the complex interactions among the constituent 
nucleons. 
Despite the complex nature of the low-energy effective forces at work
and the large number of participating particles, 
collective nuclei give rise to strikingly regular excitation spectra,
signaling the presence of underlying symmetries~\cite{Talmi}. 
The theme of ``simplicity out of complexity" 
and the understanding of simple emergent
behavior are major challenges facing the study of 
almost any many-body system, from atomic nuclei to nanoscale and 
macroscopic systems~\cite{Anderson72}.

Although, usually, a many-body Hamiltonian 
does not conform to a dynamical symmetry (DS) limit~\cite{Iac06}, 
the possibility exists that certain symmetries
are obeyed by only a subset of its eigenstates.
This situation, referred to as partial dynamical symmetry 
(PDS)~\cite{Leviatan11},
was shown to be relevant to specific 
nuclei~\cite{Leviatan11,Leviatan96,Isacker99,Leviatan02,
Escher00,Rowe01,Isacker08,Ramos09,Leviatan13} 
and molecules~\cite{Ping97}. 
In parallel, the notion of quasi dynamical symmetry (QDS)
was introduced and discussed in the context of nuclear 
models~\cite{Carvalho86,Bahri00,Rochford88,
Turner05,Rowe04,Rosensteel05,Macek09,Macek10}. 
While QDS can be defined mathematically
in terms of embedded representations~\cite{Rowe88,Rowe04b},
its physical meaning is that 
several observables associated with particular 
eigenstates, 
may be consistent with a certain symmetry which in fact is broken 
in the Hamiltonian.
This typically occurs for a Hamiltonian
transitional between two DS limits
which retains, for a certain range of its parameters,
the characteristics of one of those limits.
This ``apparent'' symmetry is due to a coherent mixing of 
representations in selected states, 
imprinting an adiabatic motion and increased 
regularity~\cite{Rosensteel05,Macek09,Macek10}. 

PDS and QDS are applicable to any many-body problem 
(bosonic and fermionic) endowed with an algebraic structure. 
They play a role in diverse phenomena 
including nuclear and molecular 
spectroscopy~\cite{Leviatan11,Leviatan96,Isacker99,Leviatan02,
Escher00,Rowe01,Isacker08,Ramos09,Leviatan13,Ping97,
Carvalho86,Bahri00,Rochford88}, quantum phase 
transitions~\cite{Turner05,Rowe04,Rosensteel05,lev07} 
and mixed regular and chaotic dynamics~\cite{Macek09,Macek10,WAL93}. 
In this Letter, a hitherto unnoticed link is established
between these two different symmetry concepts
and it is shown that coherent mixing of one symmetry (QDS)
can result in the partial conservation of a different, 
incompatible symmetry (PDS). An empirical manifestation of such a linkage 
is presented.

Algebraic models 
provide a convenient framework for exploring the role of 
symmetries~\cite{BNB}.
One such framework is the interacting boson model (IBM)~\cite{Iachello87}, 
which has been widely used to describe quadrupole 
collective states in nuclei
in terms of $N$ monopole ($s^\dag$) and quadrupole ($d^\dag$) bosons, 
representing valence nucleon pairs.
The model has $U(6)$ as a spectrum generating algebra 
and exhibits three DS limits, 
associated with chains of nested subalgebras, 
starting with $U(5)$, $O(6)$, and $SU(3)$, respectively. 
These solvable limits correspond to known benchmarks of 
the geometric description of nuclei~\cite{Bohr75}, 
involving vibrational [$U(5)$], $\gamma$-soft [$O(6)$], 
and rotational [$SU(3)$] types of dynamics.
In what follows we employ the IBM as test ground for connecting 
the PDS and QDS notions. The particular example considered,
namely, $SU(3)$ QDS as an emanation of $O(6)$ PDS,
is shown to have approximate validity in many deformed rare-earth nuclei.

One particularly successful approach within the IBM is the extended 
consistent-Q formalism (ECQF)~\cite{Warner83,Lipas85},
which is frequently used for the interpretation and classification of 
nuclear data. It uses the same quadrupole operator, 
$\hat Q^\chi=d^\dag s+s^\dag\tilde d+\chi\,(d^\dag\tilde d)^{(2)}$, 
in the $E2$ transition operator
and in the Hamiltonian, the latter being written as
\begin{equation}
\hat H_{\rm ECQF}=
\omega\left[(1-\xi)\,\hat n_d-\frac{\xi}{4N}\,
\hat Q^\chi\cdot\hat Q^\chi\right] ~,
\label{eq:Hamiltonian}
\end{equation}
where $\hat n_d$ is the d-boson number operator, 
$\hat Q^\chi\cdot\hat Q^\chi$ is the quadrupole interaction,
and the dot implies a scalar product. 
The parameters $\omega$, $\xi$, and $\chi$
are fitted to empirical data or calculated microscopically if possible;
$\xi$ and $\chi$ are the sole structural parameters of the model
since $\omega$ is a scaling factor.
The parameter ranges $0\leq\xi\leq1$ and $-\frac{\sqrt{7}}{2}\leq\chi\leq0$ 
interpolate between the $U(5)$, $O(6)$, and $SU(3)$ DS limits, 
which are reached for 
$(\xi,\chi)=(0,\chi)$, $(1,0)$, and $(1,-\frac{\sqrt{7}}{2})$, respectively.
It is customary to 
represent the parameter space by a symmetry triangle~\cite{Casten83}, 
whose vertices correspond to these limits. 
The ECQF has been used extensively
for the description of nuclear properties
(see, {\it e.g.}, Ref.~\cite{McCutchan04}) 
and it was found that rotational nuclei
are best described by ECQF parameters in the interior of the triangle,
away from the naively expected $SU(3)$ DS limit. 
The SU(3) mixing was found to be strong and coherent, {\it i.e.}, 
the same for all rotational states in a band, exemplifying 
a $SU(3)$-QDS~\cite{Rosensteel05,Macek09,Macek10}. In what follows we examine 
the $O(6)$ symmetry properties of ground-band states in such nuclei, 
in the rare-earth region, using the ECQF of the IBM. 

The $O(6)$ DS basis states are specified by 
quantum numbers $N$, $\sigma$, $\tau$, and $L$,
related to the algebras in the chain 
$U(6)\supset O(6)\supset O(5)\supset O(3)$~\cite{Arima79}. 
Given an eigenstate $\Psi$ of the ECQF Hamiltonian~(\ref{eq:Hamiltonian}),
its expansion in the O(6) basis reads
\begin{equation}
\ket{\Psi(\xi,\chi)}=
\sum_i{\alpha_i(\xi,\chi)\,\ket{N,\sigma_i,\tau_i,L}} ~,
\label{eq:o6decomp}
\end{equation}
where the sum is over all basis states
and, for simplicity, the dependence of $\Psi$ and $\alpha_i$
on the boson number $N$ and the angular momentum $L$ is suppressed. 
The degree of $O(6)$ symmetry of the state $\Psi$
is inferred from the fluctuations in $\sigma$ which can be calculated as
\begin{equation}
\Delta\sigma_\Psi=
\sqrt{\sum_i{\alpha_i^2\,\sigma_i^2} 
-\left(\sum_i{\alpha_i^2\,\sigma_i}\right)^2} ~.
\label{eq:fluc2}
\end{equation}
If $\Psi$ carries an exact $O(6)$ quantum number,
$\sigma$ fluctuations are zero, $\Delta\sigma_\Psi=0$.
If $\Psi$ contains basis states with different $O(6)$ quantum numbers,
then $\Delta\sigma_\Psi>0$, indicating that the $O(6)$ symmetry is broken. 
Note that $\Delta\sigma_\Psi$ also vanishes
for a state with a mixture of components with the same $\sigma$
but different $O(5)$ quantum numbers $\tau$,
corresponding to a $\Psi$ with good $O(6)$ but mixed $O(5)$ character.
This method of quantifying the $O(6)$ purity of states
has already been applied to $^{124}$Xe~\cite{Rainovski10}. 
Also, $\Delta\sigma_\Psi$ has the same physical content as 
wave-function entropy which, 
upon averaging over all eigenstates,
discloses the global DS content of a given Hamiltonian~\cite{Cejnar98}. 
We examine here the fluctuations $\Delta\sigma_\Psi$
for the entire parameter space of the ECQF Hamiltonian~(\ref{eq:Hamiltonian})
for values of $N$ up to 60,
using the ArbModel code~\cite{Heinze}. 

\begin{figure}[t]
\includegraphics{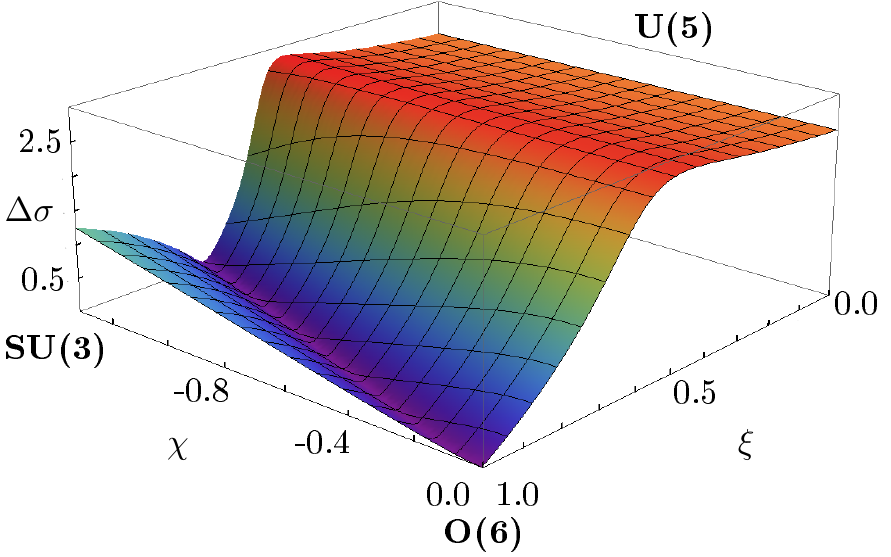}
\caption{(Color online)
Ground-state fluctuations $\Delta\sigma_{\rm gs}$~(\ref{eq:fluc2}) 
for the ECQF Hamiltonian~(\ref{eq:Hamiltonian})
with $N\!=\!14$ bosons. 
The fluctuations vanish at the $O(6)$ DS limit,
saturate towards the $U(5)$ DS limit,
and are of the order $10^{-2}$ in the valley.}
\label{fig:3d}
\end{figure}

Results of this calculation for the ground state, $\Psi=0^+_{\rm gs}$, 
with $N=14$ and parameters $\xi\in[0,1]$, $\chi\in[-\frac{\sqrt{7}}{2},0]$, 
are shown in Fig.~\ref{fig:3d}. 
At the $O(6)$ DS limit ($\xi=1$, $\chi=0$)
$\Delta\sigma_{\rm gs}$ vanishes per construction
whereas it is greater than zero for all other parameter pairs.
Towards the $U(5)$ DS limit ($\xi=0$),
the fluctuations reach a saturation value of 
$\Delta\sigma_{\rm gs}\approx 2.47$. 
At the $SU(3)$ DS limit ($\xi=1$, $\chi=-\frac{\sqrt{7}}{2}$) 
the fluctuations are $\Delta\sigma_{\rm gs}\approx 1.25$. 
In both cases the $O(6)$ symmetry is completely dissolved
as measured by $\sigma_{\rm crit}=0.849$~\cite{Rainovski10}.
Surprisingly, there is a previously unrecognized valley
of almost vanishing $\Delta\sigma_{\rm gs}$ values,
two orders of magnitude lower than at saturation. 
This region represents a parameter range of the IBM, 
outside the $O(6)$ DS limit, 
where the ground-state wave function
exhibits an exceptionally high degree of purity
with respect to the $O(6)$ quantum number $\sigma$.
\begin{figure}
\includegraphics{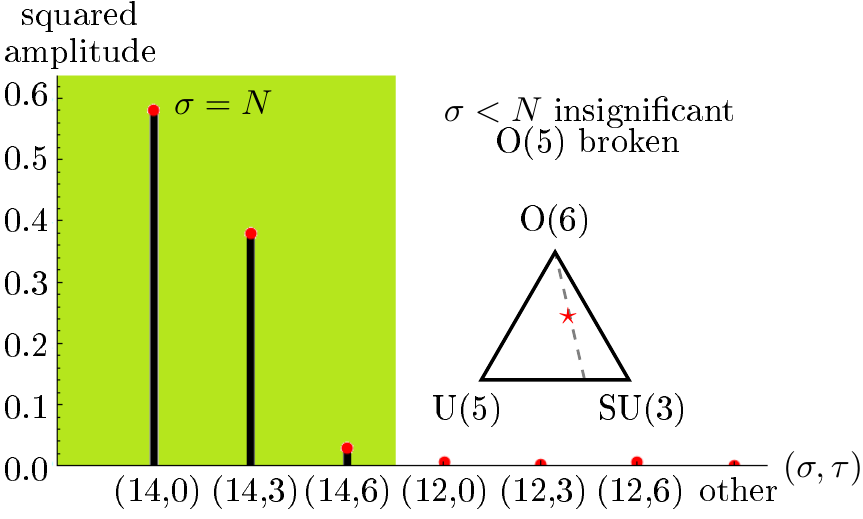}
\caption{(Color online)
Squared amplitudes $\alpha_i^2$ in the expansion~(\ref{eq:o6decomp})
of the $0^+_{\rm gs}$ ground state of the ECQF 
Hamiltonian~(\ref{eq:Hamiltonian})
for $\xi=0.84$ and $\chi=-0.53$
(indicated by the red star in the symmetry triangle
and appropriate for $^{160}$Gd).}
\label{fig:wavefunction}
\end{figure}

The ground-state wave functions in the valley of low $\Delta\sigma_{\rm gs}$ 
can be analyzed with the help of the $O(6)$ decomposition~(\ref{eq:o6decomp}). 
At the $O(6)$ DS limit only one $O(6)$ basis state, with $\sigma=N$ and 
$\tau=0$ contributes, while outside this limit 
the wave function consists of multiple $O(6)$ basis states.
Investigation of the wave function for parameter combinations inside 
the valley reveals an overwhelming dominance of the $O(6)$ basis states 
with $\sigma=N$.
This is seen in Fig.~\ref{fig:wavefunction}
for the ground-state wave function of the ECQF 
Hamiltonian~(\ref{eq:Hamiltonian})
at $\xi=0.84$ and $\chi=-0.53$ with $N=14$,
parameter values that apply to the nucleus $^{160}$Gd discussed below.
The $\sigma=N$ states comprise 
more than 99\% of the ground-state wave function
at the bottom of the valley
and their dominance causes $\Delta\sigma_{\rm gs}$ to be small. 
Furthermore, it is evident that at the same time
the $O(5)$ symmetry is broken,
as basis states with different quantum number $\tau$
contribute significantly to the wave function. 
Consequently, the valley can be identified
as an entire region in the symmetry triangle
with an approximate PDS of type III~\cite{Leviatan11},
which means that some of the eigenstates exhibit some of the symmetries.
Outside this valley the ground state is a mixture of several $\sigma$ values
and $\Delta\sigma_{\rm gs}$ increases.
In the $SU(3)$ DS limit the $\sigma=N$ components
constitute 67\% of the wave function
and in the $U(5)$ DS limit and throughout the plateau of saturated 
$\Delta\sigma_{\rm gs}$
this contribution drops below 1\%.
This region of approximate ground-state $O(6)$ symmetry  
is similar to the previously established 
``arc of regularity"~\cite{Alhassid91}
which is a region of reduced mixing inside the IBM parameter space
attributed to an approximate $SU(3)$ symmetry~\cite{Bonatsos10}.

\begin{figure}
\includegraphics{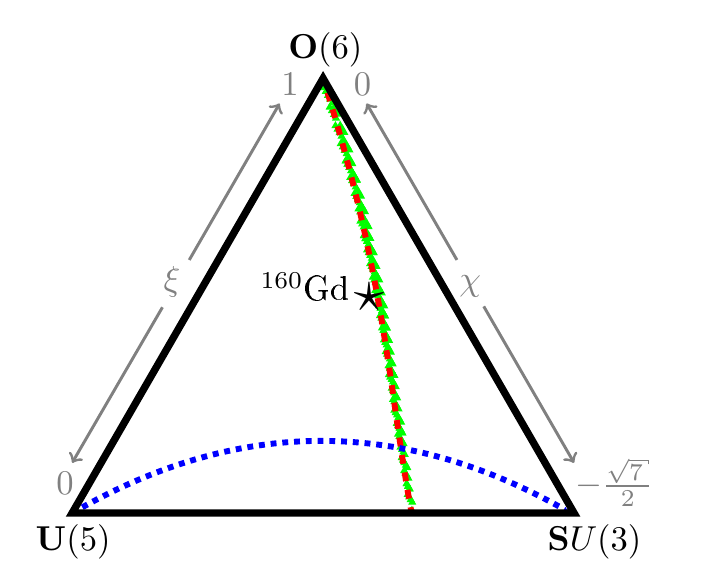}
\caption{(Color online)
The ECQF symmetry triangle with the position of 
the nucleus $^{160}$Gd indicated by a star. 
The green area shows the region of low $\Delta\sigma_{\rm gs}$,  
calculated from Eq.~(\ref{eq:fluc2}) for $N= 60$. 
The red dashed line shows the same region
of approximate ground-state $O(6)$ symmetry, 
as predicted by Eq.~(\ref{eq:Theo}) for large $N$. 
The blue dotted line shows the ``arc of regularity''~\cite{Alhassid91}.}
\label{fig:triangle}
\end{figure}

An argument for the existence
of the valley of ground-state $O(6)$ symmetry
can be given in terms of the following Hamiltonian~\cite{Leviatan02}:
\begin{eqnarray}
\hat H_{\rm M} &=&
-\hat C_{O(6)}+
\hat N(\hat N+4)+
2\alpha\hat C_{O(5)}-
\alpha\hat C_{O(3)}
\nonumber\\
&& +\,2\alpha\hat n_d(\hat N-2)
+\sqrt{14}\alpha(d^\dag s
+s^\dag \tilde d)\cdot(d^\dag\tilde d)^{(2)}\:,\qquad
\label{eq:Hpds}
\end{eqnarray}
where $\hat C_G$ denotes the quadratic Casimir operator of the group 
$G$~\cite{Iachello87},
$\hat N$ is the total boson number operator, 
and $\alpha$ is a parameter.
The Hamiltonian~(\ref{eq:Hpds}) generates a PDS of type III~\cite{Leviatan11}. 
For $\alpha= 0$, $\hat H_{\rm M}$ has exact $O(6)$ symmetry
whereas for $\alpha> 0$ the last two terms introduce $O(6)$-symmetry breaking.
However, the yrast states of this Hamiltonian,
projected from the IBM intrinsic state with 
shape variables~\cite{Bohr75}, $\beta=1$ and $\gamma=0$, 
keep exact $O(6)$ symmetry ($\sigma=N$)
but break the $O(5)$ symmetry (mixed $\tau$)
for all values of $\alpha>0$~\cite{Leviatan02}. 
Interestingly, although $\hat H_{\rm M}$ differs from $\hat H_{\rm ECQF}$, 
the overlap between their $0^+_{\rm gs}$ ground states 
maximizes (more than 99\%) in extended regions of $(\xi,\chi$) 
inside the valley of low $\Delta\sigma_{\rm gs}$. 
This suggests that the ($\beta=1,\gamma=0$) intrinsic state 
provides a good approximation, in a variational sense, 
to the ground band of $\hat H_{\rm ECQF}$ along the valley.
The equilibrium deformations for a given IBM Hamiltonian are found 
by minimizing an energy surface, $E(\beta,\gamma)$,
obtained by its expectation value in an intrinsic state 
which is a condensate of $N$ bosons, 
$b^{\dagger}_{c}\propto \beta\cos\gamma d^{\dagger}_0 
+\beta\sin\gamma(d^{\dagger}_2+d^{\dagger}_{-2})/\sqrt{2} 
+ s^{\dagger}$, that depends parametrically on 
$(\beta,\gamma)$~\cite{Ginocchio80,Dieperink80}.
Apart from a constant,
$E(\beta,\gamma)\propto(1+\beta^2)^{-2}\beta^2
\left[a-b\beta\cos{3\gamma}+c\beta^2\right]$, 
where $a$, $b$, and $c$ are coefficients depending on the Hamiltonian. 
The two extremum equations, 
$\partial E/\partial\beta=\partial E/\partial\gamma=0$, 
have $\beta=1$ and $\gamma=0$ as a solution, provided $b=2c$. 
For large $N$, the coefficients of $\hat H_{\rm ECQF}$ 
are $b=-\omega\xi\sqrt{\frac{2}{7}}\chi/N$ and 
$c=\omega\left[1-\xi-\xi\chi^2/14\right]/N$. 
Thus, in the valley of low $\Delta\sigma_{\rm gs}$ 
the desired condition, $b=2c$, fixes $\xi$ to be
\begin{equation}
\xi=\frac{1}{1-\sqrt{\frac{1}{14}}\chi+\frac{1}{14}\chi^2} ~.
\label{eq:Theo}
\end{equation}
As seen in Fig.~\ref{fig:triangle}, 
this relation predicts the location of the region of approximate 
ground-state $O(6)$ symmetry for large $N$ very precisely. 
For small $N$ its precision decreases somewhat due to finite-$N$ effects,
causing a more pronounced curvature of the region close to the 
$O(6)$ limit.
\begin{figure*}
\includegraphics{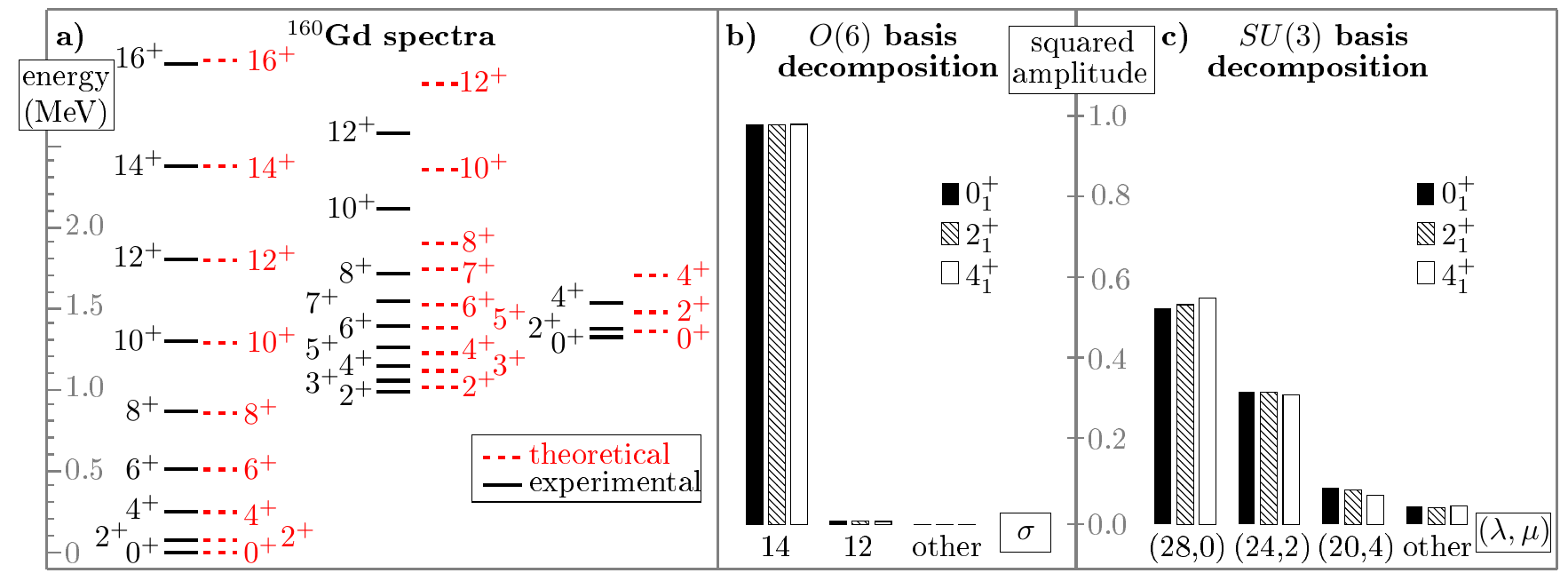}
\caption{(Color online) 
a)~The experimental spectrum of $^{160}$Gd
compared with the IBM calculation 
using the ECQF Hamiltonian~(\ref{eq:Hamiltonian})
with parameters $\xi=0.84$ and $\chi=-0.53$
taken from Ref.~\cite{McCutchan04}. 
b)~The $O(6)$ decomposition in $\sigma$ components of yrast states 
with $L=0,2,4$.
c)~The $SU(3)$ decomposition in $(\lambda,\mu)$ components of 
the same yrast states.}
\label{fig:160Gd}
\end{figure*}
\begin{table}
\caption{Calculated $\sigma$ fluctuations 
$\Delta\sigma_{L}$, Eq.~(\ref{eq:fluc2}), 
for rare earth nuclei in the vicinity of the identified region of 
approximate ground-state-$O(6)$ symmetry. 
Also shown are the fraction 
$f^{(L)}_{\rm \sigma=N}$ of $O(6)$ basis states with $\sigma = N$ contained 
in the $L\!=\!0,2,4$ states, members of the ground band. 
The structure parameters $\xi$ and $\chi$ are taken 
from \cite{McCutchan04}.\label{nuclei}}
\begin{ruledtabular}
\begin{tabular}{lccccccccc}
Nucleus & $N$ & $\xi$ & $\chi$  
& $\Delta\sigma_{0}$ & $f^{(0)}_{\rm \sigma=N}\;\;$
& $\Delta\sigma_{2}$ & $f^{(2)}_{\rm \sigma=N}\;\;$
& $\Delta\sigma_{4}$ & $f^{(4)}_{\rm \sigma=N}$\\[1mm]
\hline\\[-2mm]
$^{156}$Gd & 12 & 0.72 & -0.86 & 0.46 & 95.3\% & 
0.43 & 95.8\% & 0.38 & 96.6\% \\
$^{158}$Gd & 13 & 0.75 & -0.80 & 0.35 & 97.2\% & 
0.33 & 97.5\% & 0.30 & 97.9\% \\
$^{160}$Gd & 14 & 0.84 & -0.53 & 0.19 & 99.1\% & 
0.19 & 99.2\% & 0.17 & 99.3\% \\
$^{162}$Gd & 15 & 0.98 & -0.53 & 0.41 & 96.0\% & 
0.40 & 96.0\% & 0.40 & 96.1\% \\
$^{160}$Dy & 14 & 0.81 & -0.49 & 0.44 & 96.2\% & 
0.39 & 96.4\% & 0.36 & 96.8\% \\
$^{162}$Dy & 15 & 0.92 & -0.31 & 0.07 & 99.9\% & 
0.07 & 99.9\% & 0.06 & 99.9\% \\
$^{164}$Dy & 16 & 0.98 & -0.26 & 0.13 & 99.6\% & 
0.13 & 99.6\% & 0.13 & 99.6\% \\ 
$^{164}$Er & 14 & 0.84 & -0.37 & 0.39 & 96.5\% & 
0.37 & 96.7\% & 0.35 & 97.1\% \\
$^{166}$Er & 15 & 0.91 & -0.31 & 0.12 & 99.7\% & 
0.11 & 99.7\% & 0.10 & 99.7\% \\
\end{tabular}
\end{ruledtabular}
\end{table}

Detailed ECQF fits for energies and electromagnetic 
transitions of rare-earth nuclei, 
performed by McCutchan {\it et al.}~\cite{McCutchan04}, 
allow one to relate the structure of collective nuclei
to the parameter space of the ECQF Hamiltonian~(\ref{eq:Hamiltonian}). 
Examining the extracted ($\xi,\chi$) parameters,
one finds that several rotational nuclei in this region, 
such as $^{160}$Gd, commonly interpreted as $SU(3)$-like nuclei, 
are actually located in the valley of small $\sigma$ fluctuations. 
They can be identified as candidate nuclei
with approximate ground-state $O(6)$ symmetry.
The experimental spectrum of $^{160}$Gd,
along with its ECQF description
with $\xi=0.84$ and $\chi=-0.53$ taken from Ref.~\cite{McCutchan04},
is shown in the left panel of Fig.~\ref{fig:160Gd}. 
The middle and right panels
show the decomposition into $O(6)$ and $SU(3)$ basis states, respectively,
for yrast states with $L=0,2,4$. 
It is evident that the $SU(3)$ symmetry is broken,
as significant contributions of basis states
with different $SU(3)$ quantum numbers $(\lambda,\mu)$ occur.
It is also clear from Fig.~\ref{fig:160Gd}c
that this mixing occurs in a coherent manner
with similar patterns for the different members of the ground-state band.
This is the hallmark of a QDS~\cite{Rowe04}
and it results from the existence
of a single intrinsic wave function for the members of this band. 
On the other hand, as seen in Fig.~\ref{fig:160Gd}b, 
the yrast states with $L=0,2,4$
are almost entirely composed out of $O(6)$ basis states with $\sigma=N=14$
which implies small fluctuations $\Delta\sigma_\Psi$
and the preservation of $O(6)$ symmetry in the ground-state band.

Other rare-earth nuclei 
with ground-state bands with approximate $O(6)$ symmetry
can be identified by the same arguments. 
Their structure parameters $\xi$ and $\chi$
can be taken from Ref.~\cite{McCutchan04},
from where the fluctuations $\Delta\sigma_\Psi$
and the fractions $f_{\sigma=N}$ of squared $\sigma=N$ amplitude
can be calculated. 
Nuclei with $\Delta\sigma_{\rm gs} < 0.5$ and $f_{\sigma=N}>95\%$
are listed in Table~\ref{nuclei}.
These quantities are also calculated for yrast states 
with $L>0$ and exhibit similar values in each nucleus. 
It is evident that the IBM predicts a high degree of $O(6)$ 
purity in the ground-state-band, 
for a large set of rotational rare-earth nuclei.

These results show that the approximate $O(6)$ PDS
does hold not only for the ground state
but also for the members of the band built on top of it.
Since the entire band corresponds to a single intrinsic state,
the $SU(3)$ wave-function decomposition is similar
for the different members of the band
and therefore the notion of $SU(3)$ QDS applies. 
In addition, provided the indicated intrinsic state has 
$\beta\approx 1$ and $\gamma=0$, the notion of $O(6)$ PDS applies.
Thus a link is established between $SU(3)$ QDS and $O(6)$ PDS. 

To summarize, the method of quantum-number fluctuations
reveals the existence of a region of almost exact ground-state-band 
$O(6)$ symmetry outside the $O(6)$ DS limit of the IBM. 
The existence of a valley of small $\sigma$ fluctuations
can be understood in terms of an approximate $O(6)$ PDS of type III.
The same wave functions 
display coherent ($L$-independent) mixing of $SU(3)$ representations
and hence comply with the conditions of an $SU(3)$ QDS. 
Coherent mixing of one symmetry
may therefore result in the purity of a quantum number
associated with partial conservation of a different, incompatible symmetry.
Previously established ECQF systematics
show that many rare-earth nuclei
do exhibit these approximate 
partial $O(6)$ and quasi $SU(3)$ dynamical symmetries. 
We conclude that partial dynamical symmetries 
are more abundant than previously recognized,
may lead to coherent mixing and quasi dynamical symmetries,
and hence play a role in understanding the regular behavior of complex nuclei.
This example serves to illustrate a fundamental linkage between two 
distinct types of intermediate symmetries, PDS and QDS,
with potential implications to algebraic modeling of diverse dynamical systems.

This work has been supported by the DFG through Grant No. SFB634 
and, in part, by the Israel Science Foundation (A.L.).

\end{document}